\documentclass[pra,final,singlecolumn,showpacs,showkeys]{revtex4}
\usepackage{amsmath}
\usepackage{graphicx}
\usepackage{amsfonts}
\usepackage{amssymb}
\usepackage{subfigure}

\newcommand{\subs}[1]
{ 
	\mbox{\scriptsize{#1}}
}

\begin{document}

\title{Quantum State Synthesis of Superconducting Resonators}

\author{Roshan Sharma and Frederick W. Strauch}  \email[Electronic address: ]{Frederick.W.Strauch@williams.edu}
\affiliation{Williams College, Williamstown, MA 01267, USA}

\date{\today}

\begin{abstract}
We present a theoretical analysis of different methods to synthesize entangled states of two superconducting resonators.  These methods use experimentally demonstrated interactions of resonators with artificial atoms, and offer efficient routes to generate nonclassical states.  We analyze the theoretical structure of these algorithms and their average performance for arbitrary states and for deterministically preparing NOON states.  Using a new state synthesis algorithm, we show that NOON states can be prepared in a time linear in the desired photon number and without any state-selective interactions.
\end{abstract}

\maketitle

\section{Introduction}

In recent years we have witnessed a dramatic evolution in the quantum mechanical experiments performed with superconducting circuits.  Initially, the challenge was to fabricate, prepare, and isolate signatures of quantum behavior of the coupled motion of Cooper pairs through Josephson junctions and electrodynamic oscillations in superconducting devices \cite{Devoret89, Chiorescu04, Wallraff04, Xu2005}.  This has now become routine in the field of circuit QED \cite{Schoelkopf2008}, and the frontier is designing and manipulating the quantum states of coupled superconducting qubits (quantum bits) and resonators to achieve quantum-enhanced information processing \cite{Mariantoni11,Mariantoni11b}.  

When embarking on this new journey, the quantum mechanical engineer must decide on which degrees of freedom does she wish to manipulate: the electronic (qubit) or electromagnetic (resonator)?  There are important advantages on both sides, but, until recently \cite{Paik11}, the coherence (or quality factor) of superconducting resonators (cavities) could be significantly greater than the qubit circuits utilizing Josephson junctions.  Thus, one should consider what coherent operations can be performed with superconducting resonators as opposed to qubit circuits.  Such studies include high-fidelity measurement \cite{Johnson10, Boissonneault2010, Bishop2010}, computation \cite{Strauch11}, and error correction \cite{Nigg2013,Leghtas2013}, which all attempt to utilize the larger state space afforded by the harmonic oscillator states of a resonator to achieve greater efficiency.  

Here we consider how to efficiently manipulate these modes into desired quantum states.  In particular,  we consider theoretical methods to perform ``digital'' state synthesis of superconducting resonators, where the desired state is a superposition of Fock states \cite{Hofheinz2008,Hofheinz2009}.  An alternative ``analog'' approach uses superpositions of coherent states\cite{Leghtas2013b}.  We expect that many of the issues encountered in the digital regime will have counterparts in the analog regime, but both warrant detailed study.  In this paper, we continue the analysis of Fock state manipulations, here for the synthesis of entangled states between two resonators.

The general state synthesis problem concerns how one can prepare, with high fidelity, an arbitrarily chosen quantum state.  A {\em state synthesis algorithm} is a procedure, given a description of the target state, to identify the appropriate set of Hamiltonian controls (such as amplitudes and frequencies of control fields) that will prepare the target state from a fixed initial state.  Note that there are two senses in which the state synthesis problem is solved algorithmically.  First, a classical algorithm is typically implemented as a computer program to find the set of controls.  Second, the output of this program is itself a program, namely a sequence of operations to be applied to quantum hardware to prepare the desired state.  Thus, the state synthesis algorithms presented are a means to program future quantum machines.

\begin{figure}
\begin{center}
\includegraphics[width=4in]{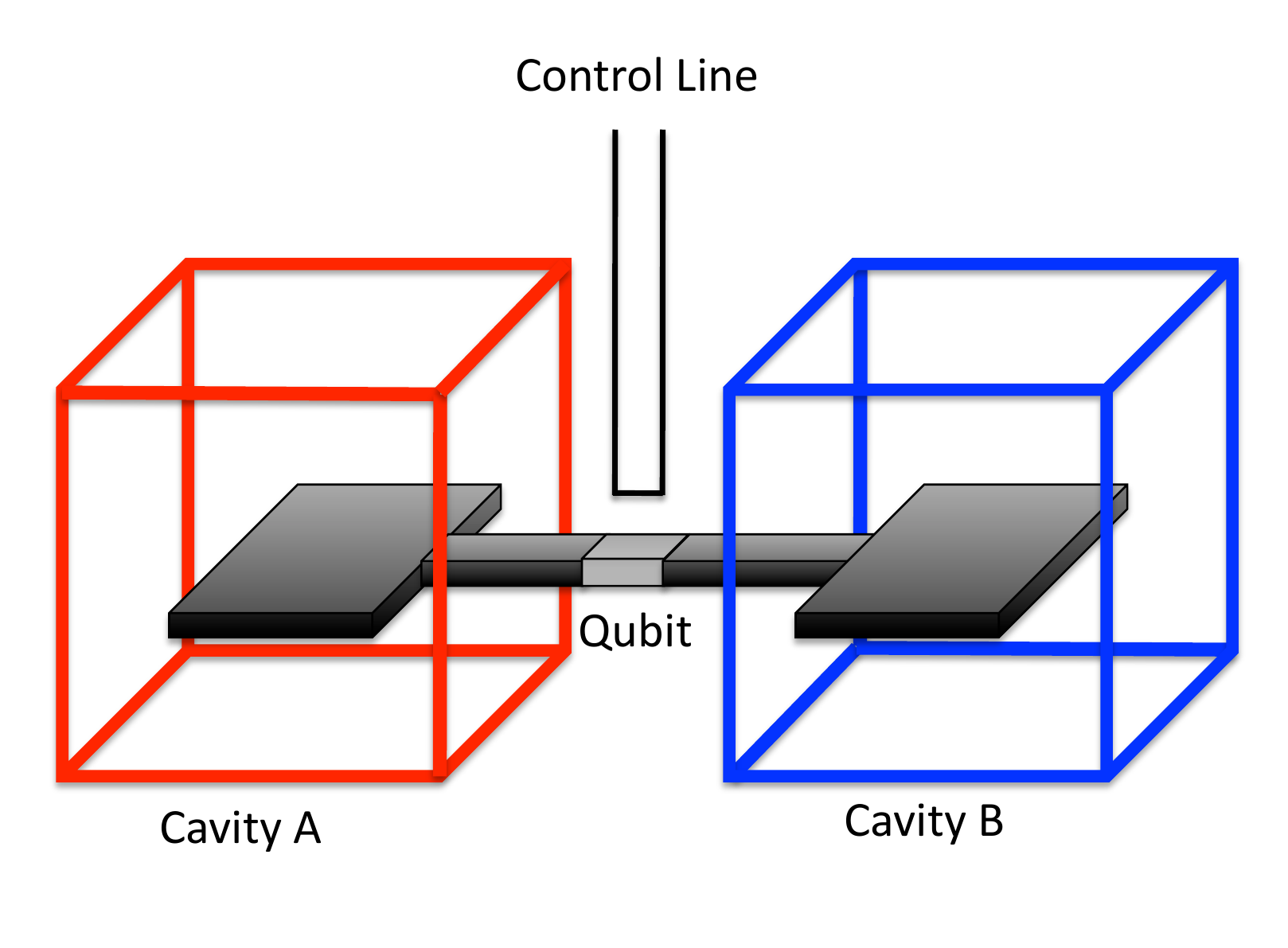}
\caption{Schematic scenario for entangled state synthesis problems, in which two cavities $A$ and $B$ are coupled by a qubit.  The qubit can be controlled to effect qubit rotations or swaps between the qubit and each cavity.}
\label{cavity_fig}
\end{center}
\end{figure}

In this paper we consider a scenario such as that depicted in Fig. 1, in which a qubit is used to couple two resonant cavities $A$ and $B$, the latter
with Fock states $|n_a \rangle \otimes |n_b\rangle$.  We will analyze algorithms that deterministically and exactly produce an arbitrarily chosen state of two resonators, of the form
\begin{equation}
|\psi_{\subs{target}} \rangle = |0\rangle_{\subs{qubit}} \otimes \sum_{n_a=0}^{N_a} \sum_{n_b=0}^{N_b} c_{n_a, n_b} |n_a\rangle \otimes |n_b \rangle.
\end{equation}
In fact, we will provide a performance analysis of two such algorithms, one a photon subtraction algorithm based on previous work \cite{Strauch2010, Strauch2012b}, and a second photon swapping algorithm new to this work.  The results obtained here can be used as a benchmark for alternative procedures to prepare such states.  These alternatives include numerical optimization methods, closed-loop control methods, or other measurement-based methods for state preparation.  Finally, we specifically consider how our algorithm compares with special-purpose NOON-state preparation \cite{Merkel10, Wang11}, and demonstrate that our new algorithm can synthesize these and a large class of entangled states without state-selective interactions.  As state-selective interactions are often weaker than direct qubit-resonator interactions, we expect these results will aid future demonstrations of entanglement in superconducting qubit-resonator systems.

This paper is organized as follows.  In Section II, we introduce the general state synthesis problem by studying how to prepare a general state of a $d$-level quantum system (i.e., a qudit).  This is followed in Section III by a presentation of the Law-Eberly algorithm for a single resonator coupled to a qubit, before addressing in Section IV the two entangled-state synthesis algorithms for two resonators coupled by a qubit.  Finally, Section V compares these algorithms for the preparation of NOON states.  We conclude in Section VI by summarizing our work and open questions.

\section{Qudit State Synthesis}

Before focusing on state synthesis problems for systems with specific Hamiltonians, it is useful to start with a simpler problem.  Thus, we begin by considering the synthesis of an arbitrary state of a $d$-level system known as a qudit (a quantum digit) \cite{Gottesman99}.  Such systems are universal for computation \cite{Stroud2000, Brennen05}, and much is known about the construction of logic \cite{Bullock05,Oleary06,Ralph07, Lanyon09} and error correction \cite{Gottesman99, Bartlett02} using such systems .  For our case, the qudit could be a nonlinear oscillator driven directly by control fields with frequencies tuned to distinct transitions, shifted by either resonant or dispersive coupling to a qubit \cite{Strauch2012}.  

Our task is to prepare the quantum state
\begin{equation}
|\psi_{\subs{target}}\rangle = \sum_{n=0}^{d-1} c_n |n\rangle
\label{qudit_state1}
\end{equation}
starting from the initial state $|0\rangle$.  For ease of analysis, we write the coefficients of the target state $|\psi\rangle$ in terms of $d$ phases and $d$-dimensional spherical polar coordinates
\begin{eqnarray}
c_0 &=& \cos \theta_0 e^{i \phi_0} \nonumber \\
c_1 &=& \sin \theta_0 \cos \theta_1 e^{i \phi_1} \nonumber \\
& \vdots & \nonumber \\
c_{d-2} &=& \sin \theta_0 \sin \theta_1 \cdots \sin \theta_{d-3} \cos \theta_{d-2} e^{i \phi_{d-2}} \nonumber \\
c_{d-1} &=& \sin \theta_0 \sin \theta_1 \cdots \sin \theta_{d-3} \sin \theta_{d-2} e^{i \phi_{d-1}},
\label{qudit_spherical}
\end{eqnarray}
and the angles have the ranges $0 \le \theta_j \le \pi/2$ and $-\pi < \phi_j \le \pi$.  Note that the first phase $\phi_0$ could
be set to zero without changing the physical problem.

The quantum state can be generated by two types of operations: the two-level rotations
\begin{equation}
\mathcal{R}_{n,n+1}(\theta) = \exp \left[ - i \frac{\theta}{2} \left( |n\rangle \langle n+1| + |n+1\rangle \langle n | \right) \right],
\end{equation}
and the single-level phase shifts
\begin{equation}
\mathcal{Z}_n(\phi) = \exp \left( i \phi |n\rangle \langle n| \right).
\end{equation}
A solution to the state synthesis problem is the following
\begin{equation}
|\psi\rangle = \mathcal{Z}_{d-1} (\phi_{d-1}) \left( \prod_{k=0}^{d-2}  \mathcal{Z}_{k}(\phi_k) \mathcal{Z}_{k+1} (\pi/2) \mathcal{R}_{k,k+1}(2 \theta_k) \right) |0\rangle,
\label{qudit_state2}
\end{equation}
where we are using a ``time-ordered'' product notation, e.g.
\begin{equation}
\prod_{j=1}^{d-1} U_j = U_{d-1} U_{d-2} \cdots U_1.
\end{equation}
For a qubit ($d=2$), this reduces to $\mathcal{Z}_0(\phi_0) \mathcal{Z}_1(\phi_1+\pi/2) \mathcal{R}_{0,1}(2 \theta_0)$, which can be combined into the product of two spin rotations (about the $x$ and $z$ axes, respectively).  This is the number of rotations required to map an arbitrary qubit state's Bloch vector from the north pole to any point on the Bloch sphere.   Similarly, this solution is a minimal approach to controlling a qudit, using a fixed set of operations for an arbitrary state of the form Eq. (\ref{qudit_state1}).  

While this solution can be verified by inspection, an alternative approach, the prototype for the state synthesis algorithms to be described below, is to find the rotations by reversing the time evolution, that is, to choose a set of operations $U_j^{\dagger}$ such that
\begin{equation}
\prod_{j=d-1}^{1} U_j^{\dagger} |\psi \rangle = U_1^{\dagger} \cdots U_{d-1}^{\dagger} |\psi \rangle = |0\rangle.
\label{qudit_product}
\end{equation} 
By simple inversion of Eq. (\ref{qudit_product}) we can use this solution to the inverse evolution equation to find a solution of the state synthesis problem given by Eq. (\ref{qudit_state2}).  The algorithmic approach is to choose each operation to ``zero out'' an amplitude of the target state.  Specifically, we index the steps of the algorithm and define the quantum state
\begin{equation}
|\psi_j \rangle = U_{j}^{\dagger} |\psi_{j+1}\rangle,
\end{equation}
where $|\psi_d\rangle = |\psi\rangle$ and $j = d-1 \to 0$.  The operator $U_{j}^{\dagger}$ is then chosen so that
\begin{equation}
\langle j | \psi_j \rangle = 0.
\end{equation}
Using the rotations specified above, we can set
\begin{equation}
U_{j}^{\dagger} = \mathcal{R}_{j-1,j}^{\dagger}  (\gamma_{j})  \mathcal{Z}_{j}^{\dagger}  (\beta_{j}) \mathcal{Z}_{j-1}^{\dagger}(\alpha_{j}), 
\label{qudit_operator}
\end{equation}
where
\begin{eqnarray}
\alpha_{j} &=&  \mbox{arg} \left( \langle j-1|\psi_{j+1} \rangle \right) \nonumber \\
\beta_{j} &=& \frac{\pi}{2} + \mbox{arg} \left( \langle j| \psi_{j+1} \rangle \right) \nonumber \\
\gamma_{j} &=& 2 \ \mbox{arctan} \left( \left\vert \frac{\langle j | \psi_{j+1} \rangle}{\langle j-1|\psi_{j+1}\rangle} \right\vert \right).
\label{qudit_angles}
\end{eqnarray}

Before verifying that this produces the same solution as Eq. (\ref{qudit_state2}), let us consider the the first rotation $U_{d-1}^{\dagger}$ in Eq. (\ref{qudit_product}) (the last rotation of the forward sequence).  This is chosen to remove the highest state $|d-1\rangle$ from the superposition in $|\psi_d\rangle$.  Using Eqs. (\ref{qudit_state1}) and (\ref{qudit_operator}) we have
\begin{equation}
\langle d-1 | U_{d-1}^{\dagger} |\psi\rangle = e^{-i \beta_{d-1}} \cos (\gamma_{d-1} /2) c_{d-1} + i e^{-i \alpha_{d-1} } \sin(\gamma_{d-1}/2) c_{d-2}.
\end{equation}
Using the spherical coordinates for $c_{d-2}$ and $c_{d-1}$ from Eq. (\ref{qudit_spherical}) (and cancelling common terms), we thus require
\begin{equation}
\cos(\gamma_{d-1}/2) \sin \theta_{d-2} e^{-i \beta_{d-1}} e^{i \phi_{d-1}} + i \sin(\gamma_{d-1}/2) \cos \theta_{d-2} e^{-i \alpha_{d-1}} e^{i \phi_{d-2}} =0
\end{equation}
which is satisfied by
\begin{eqnarray}
\alpha_{d-1} &=& \phi_{d-2} \nonumber \\
\beta_{d-1} &=& \frac{\pi}{2} + \phi_{d-1} \nonumber \\ 
\gamma_{d-1} &=& 2 \theta_{d-2}, 
\end{eqnarray}
in complete agreement with Eq. (\ref{qudit_angles}).

The same procedure works for each $U_{j}^{\dagger}$, the only difference being that the phases of $\langle j | \psi_{j+1}\rangle$ have already been set to zero for $j=d-2, d-3, \cdots$ (in the previous step), so that in general we find
\begin{eqnarray}
\alpha_{j} &=& \phi_{j-1} \nonumber \\
\beta_{j} &=& \frac{\pi}{2} + \delta_{j,d-1} \phi_j \nonumber \\
\gamma_{j} &=& 2 \theta_{j-1}.
\end{eqnarray}
Using these angles, we see that
\begin{equation}
|\psi\rangle = \prod_{j=1}^{d-1} U_j |0\rangle = U_{d-1} \cdots U_1 |0\rangle,
\end{equation}
where
\begin{equation}
U_j = \mathcal{Z}_{j-1} (\alpha_j) \mathcal{Z}_{j} (\beta_j) \mathcal{R}_{j-1,j}(\gamma_j),
\end{equation}
agrees with Eq. (\ref{qudit_state2}) after setting $k = j-1$.  Note, however, that the choice of the angles is not unique.  We could have set $\alpha_j = 0$ and $\beta_j = \pi/2 + (\phi_j - \phi_{j-1})$ to achieve the same result.  

For an arbitrary target state $|\psi\rangle$, we can characterize the performance of this algorithm in terms of the resources needed to construct the state.  These resources could be analyzed in terms of the number of controls required, the energy associated with each control, and the duration over which the control fields act.  For simplicity, we consider that each two-state rotation can occur with an effective Rabi frequency $\Omega$ and each phase shift with $\pm \Delta \omega$.  Then, this algorithm produces a set of  $d-1$ phase shifts (assuming $\alpha$ and $\beta$ can occur in parallel) and $d-1$ rotations, such that the overall time is
\begin{equation}
T =  \frac{1}{\Delta \omega} \sum_{j=1}^{d-1} |\beta_j| + \frac{1}{\Omega} \sum_{j=1}^{d-1} \gamma_{j}.
\end{equation}
The average time required can be found by averaging over the unit circle (for $\beta_j)$ and the spherical coordinates in Eq. (\ref{qudit_spherical}) (for $\gamma_j = 2\theta_j$).  We find that $\langle \beta_j \rangle = \pi/2$ and
\begin{equation}
\langle \theta_j \rangle = \frac{ \int_{0}^{\pi/2} \theta \left( \sin \theta \right)^{d-2-j} d \theta }{ \int_{0}^{\pi/2} \left( \sin \theta \right)^{d-2-j} d \theta } \approx \frac{\pi}{2} - \frac{\pi}{4} \frac{1}{\sqrt{d-2-j}}.
\end{equation}
Thus, we find that 
\begin{equation}
\langle T \rangle = \left( \frac{\pi}{\Omega} + \frac{\pi}{2 \Delta \omega} \right) (d-1) - \frac{\pi}{2 \Omega} \sum_{k=1}^{d-1} \frac{1}{\sqrt{k}}.
\end{equation}
Thus, this particular sequence takes a time that grows roughly linear in the Hilbert space dimension $d$, with timescales given by $1/\Omega$ and $1/\Delta \omega$.


\section{Law-Eberly Algorithm}

Having illustrated the properties of qudit state synthesis, we proceed to a qubit-oscillator system, appropriate for superconducting circuits and resonators.  This algorithm was first put forward by Law and Eberly in the context of cavity-QED \cite{Law96}, and experimentally demonstrated using the internal and vibrational states of a trapped ion \cite{BenKish03}.   The superconducting experiments \cite{Hofheinz2008,Hofheinz2009} demonstrated exquisite control over the combined Hilbert space of the qubit-resonator system.

Here we review this problem, namely how to synthesize an arbitrary state of harmonic oscillator mode (the resonator) by using a two-level auxiliary system (qubit).  The target state is taken to be
\begin{equation}
|\psi_{\subs{target}}\rangle = |0\rangle \otimes \sum_{n=0}^{N_{\subs{max}}} c_n |n\rangle,
\label{le_target}
\end{equation}
in which the resonator has a maximum photon number $N_{\subs{max}}$.  The systems are coupled by Jaynes-Cummings-type swapping interaction, with a Hamiltonian (in the interaction picture) of the form
\begin{equation}
\mathcal{H}/\hbar =\frac{1}{2} \Delta \omega(t) \sigma^{\dagger} \sigma + \frac{1}{2} \Omega(t) \sigma_x + g(t) \left( \sigma^{\dagger} a + \sigma  a^{\dagger} \right),
\end{equation}
where $\sigma = |0\rangle \langle 1|$ is the lowering operator for the qubit and the control fields ($\Delta \omega(t)$, $\Omega(t)$, and $g(t)$) are assumed to be under experimental control.  These control fields enable unitary operations of the form
\begin{equation}
S(\theta) = \exp \left[- i \theta \left( a \sigma^{\dagger} + a^{\dagger} \sigma \right) \right],
\end{equation}
\begin{equation}
R(\theta) = \exp \left( - i \frac{\theta}{2} \sigma_x \right),
\end{equation}
and
\begin{equation}
Z(\phi) = \exp \left( - i \frac{\phi}{2} \sigma_z \right).
\end{equation}
The Law-Eberly algorithm will be expressed in terms of these operations.  

The state-synthesis procedure follows a similar pattern as the qudit case presented above.  We first set  
\begin{equation}
|\psi_j\rangle = U_j^{\dagger} |\psi_{j+1} \rangle
\label{le_iter}
\end{equation}
where 
\begin{equation}
U_j^{\dagger} = R^{\dagger}(\gamma_j) Z^{\dagger}(\beta_j) S^{\dagger}(\theta_j) Z^{\dagger}(\alpha_j).
\label{le_operator}
\end{equation}
and $|\psi_{N+1} \rangle = |\psi_{\subs{target}}\rangle$.  Here $\alpha,\beta,\gamma, \ \mbox{and} \ \theta$ are chosen so that at each step
\begin{equation}
\langle 0, j|\psi_j \rangle = \langle 1,j | \psi_j \rangle = 0,
\end{equation}
These angles are then found for each $j = N \to 1$, after which $|\psi_1 \rangle = |0,0\rangle$.  The inverse sequence specifies how to prepare the target state using only qubit rotations, phase shifts, or qubit-resonator swaps.

To see how this can be accomplished, it is convenient to break Eq. (\ref{le_iter}) into two steps by defining
\begin{equation}
|\psi_{j+1/2} \rangle = S^{\dagger}(\theta_j) Z^{\dagger} (\alpha_j) |\psi_{j+1}\rangle,
\label{le_iter1}
\end{equation}
and
\begin{equation}
|\psi_j \rangle = R^{\dagger}(\gamma_j) Z^{\dagger}(\beta_j) |\psi_{j+1/2} \rangle.
\label{le_iter2}
\end{equation}
For convenience, we also define
\begin{equation}
\psi_{q,k}(j) = \langle q,k | \psi_j \rangle, \ \mbox{where} \ q = 0 \ \mbox{or} \ 1.
\end{equation}

The first step solves $\psi_{0,j}(j+1/2) = 0$.  Using Eq. (\ref{le_iter1}), this reduces to
\begin{equation}
e^{i \alpha_j/2} \cos \left( \sqrt{j} \theta_j \right) \psi_{0,j}(j+1) + i e^{-i \alpha_j/2} \sin \left(\sqrt{j} \theta_j \right) \psi_{1,j-1}(j+1) = 0
\end{equation}
or
\begin{equation}
e^{-i \alpha_j} \tan \left( \sqrt{j} \theta_j \right) = i \frac{ \psi_{0,j}(j+1)}{\psi_{1,j-1}(j+1)}.
\end{equation}
This is solved by
\begin{eqnarray}
 \alpha_j &=& \mbox{arg} \left( \frac{\langle 1,j-1 | \psi_{j+1} \rangle}{ i \langle 0,j | \psi_{j+1} \rangle} \right) \nonumber \\
 \theta_j &=& \frac{1}{\sqrt{j}} \arctan \left( \left\vert \frac{ \langle 0,j| \psi_{j+1} \rangle}{\langle 1,j-1 | \psi_{j+1} \rangle} \right\vert \right).
 \label{le_angles1}
\end{eqnarray}

The second step solves $\psi_{1,j-1}(j) = 0$.  Using Eq. (\ref{le_iter2}), this reduces to
\begin{equation}
e^{-i \beta_j/2} \cos \left( \frac{\gamma_j}{2} \right) \psi_{1,j-1}(j+1/2) + i e^{i \beta_j/2} \sin \left( \frac{\gamma_j}{2} \right) \psi_{0,j-1}(j+1/2)
\end{equation}
or
\begin{equation}
e^{i \beta_j} \tan \left( \frac{\gamma_j}{2} \right) = i \frac{\psi_{1,j-1}(j+1/2)}{\psi_{0,j-1}(j+1/2)}.
\end{equation}
This, in turn, has the solution
\begin{eqnarray}
\beta_j &=& \mbox{arg} \left( \frac{i \langle 1,j-1| \psi_{j+1/2} \rangle}{\langle 0,j-1 | \psi_{j+1/2} \rangle} \right) \nonumber \\
\gamma_j &=& 2 \arctan \left( \left\vert \frac{ \langle 1,j-1| \psi_{j+1/2} \rangle }{\langle 0,j-1 | \psi_{j+1/2} \rangle } \right\vert \right).
\label{le_angles2}
\end{eqnarray}

By solving these equations for $\alpha_j, \beta_j, \gamma_j, \ \mbox{and} \ \theta_j$ for $j = N \to 0$, keeping track of $|\psi_j\rangle$ at each step, the amplitude is forced down to smaller and smaller photon numbers, so that $| \psi_0 \rangle = |0, 0 \rangle$.  The form of the sequence was chosen specifically to not send amplitude to higher photon numbers.  That is, the algorithm actually produces operators $U_j$ and states $|\psi_j\rangle$ that satisfy the condition
\begin{equation}
\langle 0, k |\psi_j\rangle = \langle 1,k |\psi_j\rangle = 0 \ \mbox{for} \ k \ge j.
\end{equation} 
This condition is the most challenging to generalize to more resonators.  

\begin{figure}
\begin{center}
\includegraphics[width=6in]{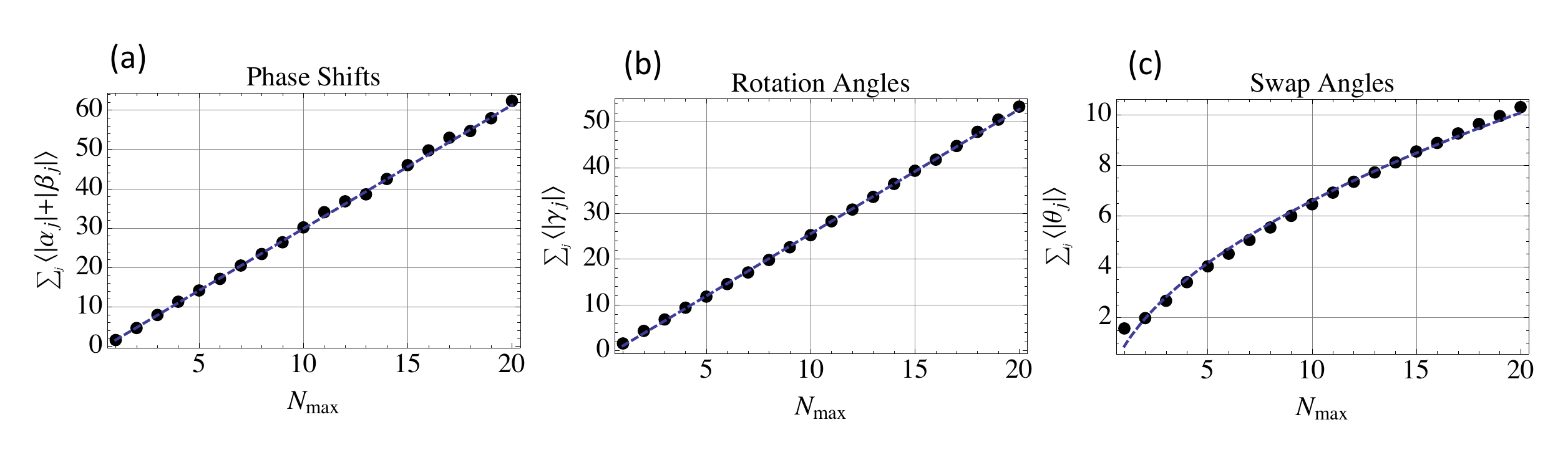}
\caption{Averaged total angles for the state synthesis sequence using the Law-Eberly algorithm.}
\label{law_eberly_fig}
\end{center}
\end{figure}

The average values of  $\alpha_j, \beta_j, \gamma_j, \ \mbox{and} \ \theta_j$ can be used to find the average time required, assuming constant controls $\pm \Delta \omega$, $\Omega$ and $g$:
\begin{equation}
T =  \frac{1}{\Delta \omega} \sum_{j=1}^{N_{\subs{max}}} \left( |\alpha_j| + |\beta_j| \right) + \frac{1}{\Omega} \sum_{j=1}^{N_{\subs{max}}} \gamma_{j} + \frac{1}{g} \sum_{j=1}^{N_{\subs{max}}} \theta_j.
\label{le_time}
\end{equation}
We illustrate the average angles obtained with Eqs. (\ref{le_angles1}) and (\ref{le_angles2}) in Fig. \ref{law_eberly_fig}.  Here we have generated one hundred random target states for each value of $N_{\subs{max}}$ and averaged the total of the angles used in the Law-Eberly algorithm.  Also shown are the approximations
\begin{eqnarray}
\sum_j \langle |\alpha_j| + |\beta_j| \rangle &\approx& \pi \left( N_{\subs{max}} - \frac{1}{2} \right), \nonumber \\
\sum_j \langle \gamma_j \rangle &\approx & 2.72 N_{\subs{max}} - 1.66, \nonumber \\
\sum_j \langle \theta_j \rangle &\approx& 2.65 \sqrt{N_{\subs{max}}} - 1.78,
\end{eqnarray}
obtained by fitting the numerical data.  

The linear increase of the phases and qubit rotations are expected, as each step requires such a rotation, while the square-root dependence of $\sum_j \langle \theta_j \rangle$ is due to the $\sqrt{n}$-coupling between the qubit and the $n$-photon state of the resonator.  When these averaged angles are substituted into Eq. (\ref{le_time}), we see that, just as the qudit case, an arbitrary state of the form Eq. (\ref{le_target}) can be synthesized in a time proportional to the effective Hilbert-space dimension.

\section{Two-Resonator Algorithms}

The state synthesis problem can be extended to any number of resonators, but explicit algorithms are a challenge to specify.  Early work utilized special interactions \cite{Gardiner97, Steinbach97,  Drobny98, Kneer98, Zheng2000} to enable the transfer of excitations between resonators and multi-level atoms.  These interactions, while natural to trapped-ion systems, are not directly applicable to the Hamiltonians considered here.  There have been a number of recent studies of interesting interactions that can be generated between superconducting or nanomechanical resonators \cite{ Xue07b, Mariantoni08, Semiao09, Kumar10, Sharypov12}.  As we are interested in Fock-state control, we consider the simplest system of a single qubit coupling two resonators, with each interaction of the Jaynes-Cummings form.  We further consider  algorithms that accomplish the synthesis of an {\em arbitrary} two-resonator state
\begin{equation}
|\psi_{\subs{target}} \rangle = |0\rangle \otimes \sum_{n_a=0}^{N_a} \sum_{n_b=0}^{N_b} c_{n_a, n_b} |n_a\rangle \otimes |n_b \rangle.
\label{tr_target}
\end{equation}
In general, such a state will be entangled, thus we call this the entangled-state synthesis problem.  

The interactions used in these algorithms are all based on the underlying Hamiltonian: 
\begin{equation}
\mathcal{H}/\hbar = \frac{1}{2} \Delta \omega(t) \sigma^{\dagger} \sigma + \frac{1}{2} \Omega(t) \sigma_x + g_a(t) \left( \sigma^{\dagger} a + \sigma a^{\dagger} \right) + g_b(t) \left( \sigma^{\dagger} b + \sigma b^{\dagger} \right).
\end{equation}
For our purposes, we again assume that the control fields ($\Delta \omega(t)$, $\Omega(t)$, $g_a(t)$, and $g_b(t)$) can be turned on and off at will, and thus we restrict our attention to swap operators
\begin{eqnarray}
A(\theta) &=& \exp \left[ -i \theta \left( a \sigma^{\dagger} + a^{\dagger} \sigma \right) \right], \nonumber \\
B(\theta) &=& \exp \left[ -i \theta \left( b \sigma^{\dagger} + b^{\dagger} \sigma \right) \right],
\end{eqnarray}
the single-qubit phase rotations
\begin{equation}
Z(\phi) = \exp \left( -i \frac{\phi}{2} \sigma_z \right),
\end{equation}
and the number-state-selective qubit rotations
\begin{equation}
R_{n_a,n_b}(\theta) = \exp \left( -i \frac{\theta}{2} \sigma_x \otimes |n_a, n_b\rangle \langle n_a, n_b | \right).
\end{equation}

The last operation utilizes the Stark-shift of each resonator on the qubit, and can (in principle) be extended to many resonators.  The actual operation may or may not be completely selective on an individual Fock state $|n_a,n_b\rangle$, but the basic conditions required can be satisfied provided there is some selectivity.  Such selective operations were first observed in circuit QED as number splitting \cite{Schuster07} and later used for photon measurement \cite{Johnson10} and theoretically proposed for state synthesis \cite{Strauch2010}.  The detailed physics of such interactions were further analyzed for both qudit operations \cite{Strauch11} and for state synthesis \cite{Strauch2012b}.  We will assume complete selectively here, but will discuss how entangled states can be synthesized with reduced selectivity in the next section.  

We will be using a Fock-state diagram, such as Fig. \ref{algorithm1_fock}, in which a state $|q,n_a,n_b\rangle$ with $n_a$ excitations in mode $A$, $n_b$ excitations in mode $B$, and qubit state $q$ is indicated by the node at location $(n_a, n_b)$ and internal level $q$.  Each of the operations described above corresponds to a transition between sets of states in this diagram, and the state sythesis sequence can be interpreted using paths in this diagram.  Two algorithms, to be described below, can be visualized using these diagrams.  The first algorithm, which we call the photon substraction algorithm, uses vertical and horizontal paths from top-to-bottom and left-to-right in the Fock-state diagram.  Thee second algorithm, which we call the photon swapping algorithm, uses diagonal paths from the upper-left to the lower-right.  These are the two natural choices for how to navigate the Fock-state diagram in order to program the quantum system into any desired state.  In this section, we will analyze each algorithm in detail, and compare their average performance when preparing an arbitrary two-resonator state of the form Eq. (\ref{tr_target}).  

\subsection{Algorithm 1: Photon Subtraction}

\begin{figure}
\begin{center}
\includegraphics[width=3.5in]{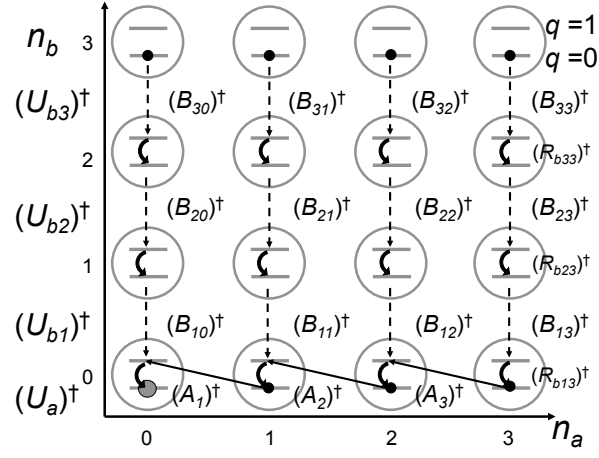}
\caption{Illustration of the state synthesis sequence using the photon subtraction algorithm.}
\label{algorithm1_fock}
\end{center}
\end{figure}

The first algorithm for superconducting resonators \cite{Strauch2010} used a strategy similar to the trapped-ion proposal by Kneer and Law \cite{Kneer98}, and involves repeated subtraction of photons from one of the resonators.  Using the Fock state diagram presented in Fig. \ref{algorithm1_fock}, state amplitudes are cleared column-by-column, row-by-row, until all of the remaining photons are in one mode only.  The final steps remove these photons by the Law-Eberly protocol described above.  

The essential steps can be written as
\begin{equation}
U = \left( \prod_{j=1}^{N_b} U_{b,j} \right) U_a
\end{equation}
where
\begin{equation}
U_{a} = \prod_{j=1}^{N_a} Z(\alpha_j) A(\theta_j) Z(\beta_j) R(\gamma_j)
\end{equation}
and
\begin{equation}
U_{b,j} = \prod_{k=0}^{N_b} Z(\alpha_{jk}) B(\theta_{jk}) Z(\beta_{jk}) R_{n_a=k}(\gamma_{jk})
\end{equation}
Read in reverse, the elements of $U_{a}$ and $U_{b,j}$ are all of the form of Eq. (\ref{le_operator}), with phases and angles calculated using the same method.  In more detail, $U_{b,j}^{\dagger}$ is a product of operations that subtract a photon from state $|0,k,j\rangle$ (transferring its amplitude to $|1,k-1,j\rangle$ and then to $|0,k-1,j\rangle$), first for $k=N_a \to 1$ (column-by-column), which is then repeated for $j=N_b \to 1$ (row-by-row).  After all of the amplitudes have been transferred to the states $|0,k,0\rangle$, $U_a$ removes these much as the original Law-Eberly algorithm.  A graphical representation of this sequence is shown in Fig. \ref{algorithm1_fock}.

To prevent amplitudes from returning to previously cleared states, it was proposed \cite{Strauch2010} to make the qubit rotations in $U_{b,j}$ number-state selective.  This could be achieved by choosing a rotation for state $(n_a = k, n_b=j-1)$ only, but the main requirement is that previously removed states with $n_b=j$ and $n_a<k$ are unaffected.  Note also that, for certain types of number-state-selective interactions, the column ordering may need to be reversed, as discussed in \cite{Strauch2012b}.  

The actual steps involved in this algorithm are nearly identical to those in the Law-Eberly algorithm.  The main challenge is to keep track of the various quantum states and the ordering of the operations.  For completeness, we include an explicit treatment here, first breaking up the quantum evolution into two stages (for the $B$ and $A$ swaps, respectively).  For the first stage, we define
\begin{eqnarray}
|\psi_{j,k+1/2} \rangle &=& B^{\dagger}(\theta_{jk}) Z^{\dagger}(\alpha_{jk}) |\psi_{j,k+1}\rangle \nonumber \\
|\psi_{j,k} \rangle &=& R_{n_a=k}^{\dagger}(\gamma_{jk}) Z^{\dagger}(\beta_{jk}) |\psi_{j,k+1/2}\rangle
\end{eqnarray}
where $k$ is the ``fast'' index (ranging from $N_a \to 0$) and $j$ is the ``slow'' index (ranging from $N_b \to 1$).  These states have the boundary conditions $|\psi_{j,N_a+1}\rangle = |\psi_{j+1,0}\rangle$ and $|\psi_{N_b, N_a+1}\rangle = |\psi_{\subs{target}}\rangle$.  Following a procedure similar to the previous section, we find
\begin{eqnarray}
\alpha_{jk} &=& \mbox{arg} \left( \frac{ \langle 1,k,j-1 | \psi_{j,k+1} \rangle }{i \langle 0, k, j| \psi_{j,k+1} \rangle }\right) \nonumber \\
\theta_{jk} &=& \frac{1}{\sqrt{j}} \arctan \left( \left\vert  \frac{\langle 0, k, j | \psi_{j,k+1} \rangle}{\langle 1, k, j-1 | \psi_{j, k+1} \rangle } \right \vert \right) \nonumber \\
\beta_{jk} &=& \mbox{arg} \left( \frac{ i \langle 1, k, j-1 | \psi_{j, k+1/2} \rangle}{ \langle 0, k, j-1 | \psi_{j, k+1/2} \rangle} \right) \nonumber \\
\gamma_{jk} &=& 2 \arctan \left( \left\vert \frac{ \langle 1, k, j-1 | \psi_{j, k+1/2} \rangle}{\langle 0, k, j-1 | \psi_{j,k+1/2} \rangle} \right\vert \right)
\label{psub_angles1}
\end{eqnarray}
These equations can be solved for $k = N_a \to 0$, $j = N_b \to 1$, until we reach the second stage.

For stage two, we define
\begin{eqnarray}
|\psi_{j+1/2}\rangle &=& A^{\dagger}(\theta_j) Z^{\dagger}(\alpha_j) |\psi_{j+1}\rangle, \nonumber \\
|\psi_j\rangle &=& R^{\dagger}(\gamma_j) Z^{\dagger}(\beta_j) |\psi_{j+1/2}\rangle, 
\end{eqnarray}
with $j$ ranging from $N_a \to 1$ and $|\psi_{N_a+1}\rangle = |\psi_{1,1} \rangle$ (the final state from stage 1).  The remaining parameters are then found by
\begin{eqnarray}
 \alpha_j &=& \mbox{arg} \left( \frac{\langle 1,j-1,0 | \psi_{j+1} \rangle}{ i \langle 0,j,0 | \psi_{j+1} \rangle} \right) \nonumber \\
\theta_j &=& \frac{1}{\sqrt{j}} \arctan \left( \left\vert \frac{ \langle 0,j,0| \psi_{j+1} \rangle}{\langle 1,j-1,0 | \psi_{j+1} \rangle} \right\vert \right) \nonumber \\
\beta_j &=& \mbox{arg} \left( \frac{i \langle 1,j-1,0| \psi_{j+1/2} \rangle}{\langle 0,j-1,0 | \psi_{j+1/2} \rangle} \right) \nonumber \\
\gamma_j &=& 2 \arctan \left( \left\vert \frac{ \langle 1,j-1,0| \psi_{j+1/2} \rangle }{\langle 0,j-1,0 | \psi_{j+1/2} \rangle } \right\vert \right).
\label{psub_angles2}
\end{eqnarray}

\begin{figure}
\begin{center}
\includegraphics[width=6in]{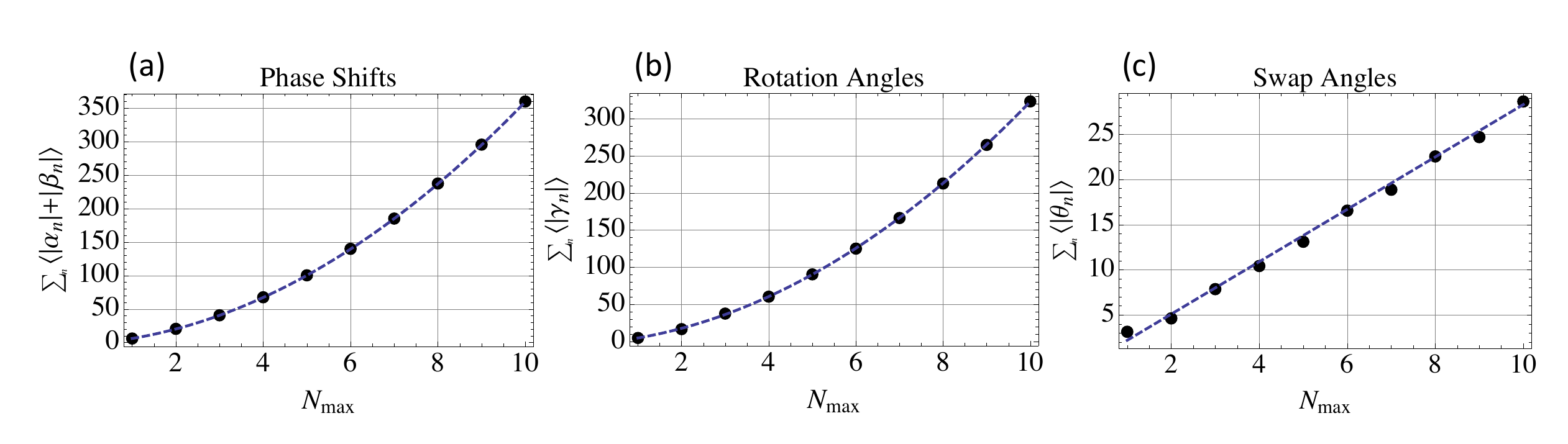}
\caption{Averaged total angles for the state synthesis sequence using the photon subtraction algorithm.}
\label{algorithm1_fig}
\end{center}
\end{figure}

The total number of operations amounts to $N_a + N_b + N_a N_b$ swaps, $N_a + N_a + N_a N_b$ rotations, and $2 (N_a + N_b + N_a N_b)$ phase shifts.  Assuming we can turn the various Hamiltonians on and off with rates $\pm \Delta \omega$, $g$, and $\Omega$ (for the phase, swap, and rotation operators, respectively), the total time for this sequence is
\begin{eqnarray}
T &=& \frac{1}{\Delta \omega} \left( \sum_{j} \left(|\alpha_j| + |\beta_j| \right) + \sum_{jk} \left( |\alpha_{jk}| + |\beta_{jk}| \right) \right) \nonumber \\
& & + \frac{1}{\Omega} \left( \sum_{j} \gamma_j + \sum_{jk} \gamma_{jk} \right) + \frac{1}{g} \left( \sum_{j} \theta_j + \sum_{jk} \theta_{jk} \right)
\end{eqnarray}
The averaged total angles are shown in Fig. \ref{algorithm1_fig}.  These were again formed by generating one hundred random target states of the form Eq. (\ref{tr_target}) with $N_a = N_b = N_{\subs{max}}$ and summing and averaging the angles produced by Eqs. (\ref{psub_angles1}) and (\ref{psub_angles2}).  Also shown are the approximations
\begin{eqnarray}
\sum_{n} \langle |\alpha_n| + |\beta_n| \rangle &\approx& \pi \left( N_{\subs{max}}^2 + \frac{3}{2} N_{\subs{max}} - \frac{1}{2} \right), \nonumber \\
\sum_n \langle \gamma_n \rangle &\approx& 2.8 N_{\subs{max}}^2 + 4.6 N_{\subs{max}} - 2.7 \nonumber, \\
\sum_n \langle \theta_n \rangle &\approx& 2.9 N_{\subs{max}} - 0.7,
\end{eqnarray}
where the sums are over all of the indices $(j,k)$ and $j$ of stages one and two, respectively.  The quadratic growth of the phase shifts and rotations match the total number of operations, while the linear growth of the swap angles is reduced by a square-root, similar to the Law-Eberly results above.  We observe that the total time required to produce an arbitrary target state again scales with effective Hilbert-space dimension.  

\subsection{Algorithm 2: Photon Swapping }

An important fact regarding state synthesis is that the solution need not be unique.  There are an infinite number of solutions, and finding the optimal solution, under constraints on time, energy, or complexity, is a hard problem.  For the algorithm just presented, we observe that each qubit rotation can add or remove one quantum of energy, which occurs $N_{\subs{max}}^2$ times.  The desired state, however, has a maximum energy of $2 N_{\subs{max}}$ (for the resonator state $|N_{\subs{max}}\rangle \otimes |N_{\subs{max}}\rangle$).  Thus, we might expect an optimal solution would have a smaller number of qubit rotations.  We will provide such a solution in this section.  For convenience, we will let the target state be a superposition of all states with quantum  number $n_a + n_b \le 2 N_{\subs{max}}$:
\begin{equation}
|\psi\rangle = |0\rangle \otimes \sum_{n_a + n_b \le 2 N_{\subs{max}}} c_{n_a,n_b} |n_a \rangle \otimes |n_b \rangle.
\end{equation}

\begin{figure}
\begin{center}
\includegraphics[width=4in]{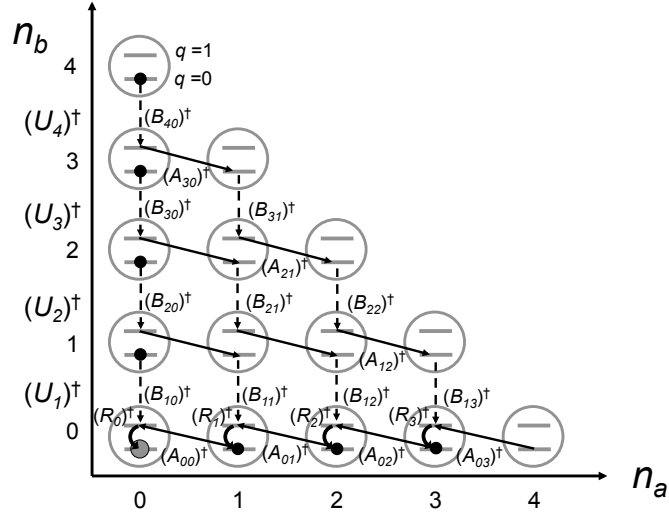}
\caption{Illustration of the state synthesis sequence using the photon swapping algorithm.}
\label{algorithm2_fock}
\end{center}
\end{figure}

The photon subtraction algorithm attempts to remove energy for each and every possible state.  However, one can just as easily move about the Fock state diagram by swapping photons between the resonators, either directly or through the qubit \cite{Mariantoni11b}.  Using the latter, we can swap photons along diagonal paths with a fixed number of quanta ($n_a + n_b + q$).  By first swapping all of the photons from resonator $b$ to $a$, we can then remove one photon at a time, using only $2 N_{\subs{max}}$ qubit rotations.  Specifically, our new algorithm is
\begin{equation}
U = \prod_{\ell = 1}^{2 N_{\subs{max}}} U_{\ell},
\end{equation}
where
\begin{equation}
U_{\ell}^{\dagger} = R_{n_a=\ell-1}^{\dagger}(\gamma_{\ell}) Z^{\dagger}(\phi_{\ell}) \prod_{m=\ell}^1 A^{\dagger}(\theta_{m-1,\ell-m}) Z^{\dagger}(\beta_{m-1,\ell-m})B^{\dagger}(\eta_{m,\ell-m})Z^{\dagger}(\alpha_{m,\ell-m})
\end{equation}
The interpretation of each operation is analogous to Algorithm 1, however, the sequence of operations is significantly different.  A graphical illustration of this photon swapping algorithm is presented in Fig. \ref{algorithm2_fock}.

In this algorithm, as we move along the diagonal path with $n_a+n_b+q = \ell$, the operator $B^{\dagger}(\eta_{m,\ell-m})$ implements the transition $|0,\ell-m,m\rangle \to |1,\ell-m,m-1\rangle$, while for $m >1$ $A^{\dagger}(\theta_{m-1,\ell-m})$ implements the transition $|1,\ell-m,m-1\rangle \to |0,\ell-m+1,m-1\rangle$.  This sequence has the effect of repeatedly swapping quanta from mode $B$ to mode $A$, until we reach $|1,\ell-1,0\rangle$.  At this point, $\theta_{0,\ell-1}$ is chosen to complete the swapping transition $|0,\ell,0 \rangle \to |1,\ell-1,0\rangle$, which is finally rotated to $|0,\ell-1,0\rangle$.  This last step need only be selective on $n_a = \ell-1$, and is so indicated in $U_{\ell}^{\dagger}$.  

For completeness, we present the detailed steps of the algorithm.  We again break each step in two by defining
\begin{eqnarray}
|\psi_{\ell,m+1/2} \rangle &=& B^{\dagger}(\eta_{m,\ell-m}) Z^{\dagger}(\alpha_{m,\ell-m}) |\psi_{\ell,m+1} \rangle \nonumber \\
|\psi_{\ell,m} \rangle &=& A^{\dagger}(\theta_{m-1,\ell-m}) Z^{\dagger}(\beta_{m-1,\ell-m}) |\psi_{\ell,m_1/2} \rangle,
\end{eqnarray}
and the various angles are calculated by the following equations:
\begin{eqnarray}
\alpha_{m,\ell-m} &=& \mbox{arg} \left( \frac{ \langle 1,\ell-m,m-1 | \psi_{\ell,m+1} \rangle }{i \langle 0, \ell-m, m| \psi_{\ell,m+1} \rangle }\right) \nonumber \\
\eta_{m,\ell-m} &=& \frac{1}{\sqrt{m}} \arctan \left( \left\vert  \frac{\langle 0, \ell-m, m | \psi_{\ell,m+1} \rangle}{\langle 1, \ell-m, m-1 | \psi_{\ell, m+1} \rangle } \right \vert \right) \nonumber \\
\beta_{m-1,\ell-m} &=& \mbox{arg} \left( \frac{ i \langle 1, \ell-m, m-1 | \psi_{\ell, m+1/2} \rangle}{ \langle 0,\ell-m+1 , m-1 | \psi_{\ell, m+1/2} \rangle} \right) \nonumber \\
\theta_{m-1,\ell-m} &=& \frac{1}{\sqrt{\ell-m+1}} \arctan \left( \left\vert \frac{\langle 1, \ell-m,m-1| \psi_{\ell,m+1/2}\rangle }{\langle 0, \ell-m+1,m-1| \psi_{\ell,m+1/2} \rangle} \right\vert \right)
\label{pswap_angles1}
\end{eqnarray}
These can be solved from $m=\ell$ until $m = 1$, for which we must modify our equations by
\begin{eqnarray}
\beta_{0,\ell-1} &=& \mbox{arg} \left( \frac{ \langle 1, \ell-1, 0 | \psi_{\ell, 1+1/2} \rangle}{ i \langle 0,\ell, 0 | \psi_{\ell, 1+1/2} \rangle} \right) \nonumber \\
\theta_{0,\ell-1} &=& \frac{1}{\sqrt{\ell}} \arctan \left( \left\vert \frac{\langle 0, \ell,0| \psi_{\ell,1+1/2}\rangle }{\langle 1, \ell-1,0| \psi_{\ell,1+1/2}\rangle } \right\vert \right).
\end{eqnarray}
This still leaves a final phase and amplitude rotation, the latter selective on $n_a=\ell-1$, with parameters
\begin{eqnarray}
\phi_{\ell} &=& \mbox{arg} \left( \frac{ i \langle 1, \ell-1, 0 | \psi_{\ell, 1} \rangle}{ \langle 0,\ell-1, 0 | \psi_{\ell, 1} \rangle} \right) \nonumber \\
\gamma_{\ell} &=& 2 \arctan \left( \left \vert \frac{\langle 1, \ell-1, 0 | \psi_{\ell, 1} \rangle}{ \langle 0,\ell-1, 0 | \psi_{\ell, 1} \rangle} \right \vert \right).
\end{eqnarray}
This sequence is then repeated for the next diagonal with $n_a+n_b+q=\ell-1$, starting with $m=\ell-1$ and $|\psi_{\ell-1,\ell} \rangle = |\psi_{\ell,1}\rangle$, and again for $\ell = 2 N_{\max} \to 1$.   

\begin{figure}
\begin{center}
\includegraphics[width=6in]{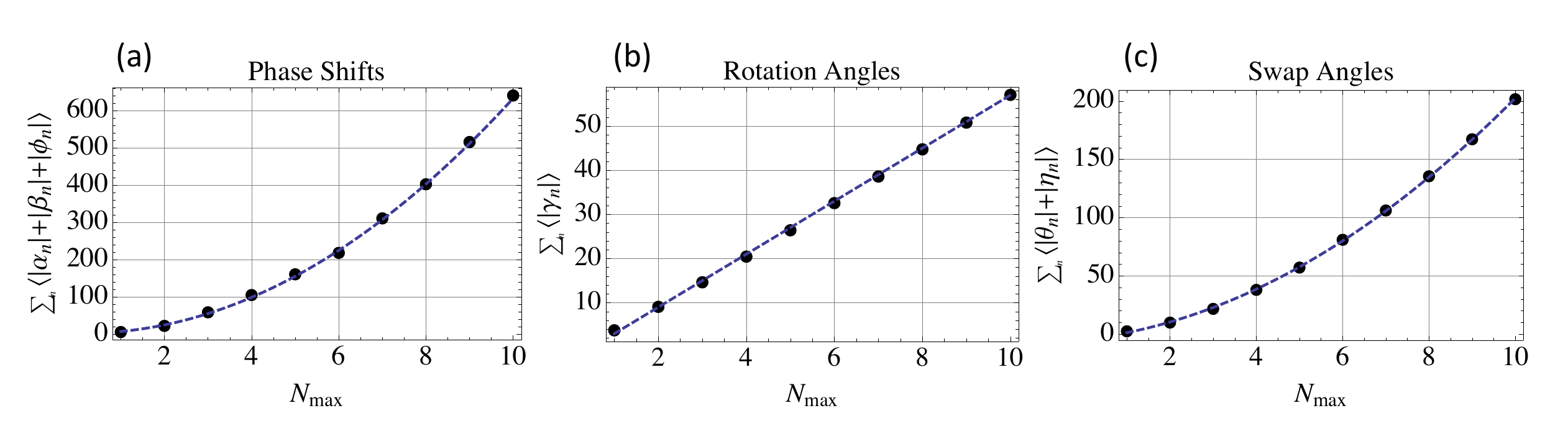}
\caption{Averaged total angles for the state synthesis sequence using the photon swapping algorithm.}
\label{algorithm2_fig}
\end{center}
\end{figure}

The expectation values for the sum of the angles, when averaged over many target states, are shown in Fig. \ref{algorithm2_fig}, along with the approximate forms
\begin{eqnarray}
\sum_n \langle |\alpha_n| + |\beta_n| + |\phi_n| \rangle &\approx& 6.4 N_{\subs{max}}^2, \nonumber \\
\sum_n \langle \gamma_n \rangle &\approx& 6 N_{\subs{max}} -3, \nonumber \\
\sum_n \langle \theta_n + \eta_n \rangle &\approx& 1.65 N_{\subs{max}}^2 + 4.2 N_{\subs{max}} - 4.5.
\end{eqnarray}
Here we see that the total rotation angles $\sum_n \langle \gamma_n \rangle$ is now {\em linear} with the maximum photon number, at the cost of an increased number of swaps (and phase rotations).  However, this can represent a significant advantage, as the qubit rotations have (so far) been required to be number-state-selective, and thus limited in Rabi amplitude $\Omega$ \cite{Strauch2012b}.  By reducing the number of such rotations, the total time can be reduced.  The overall scaling of the time, however, is again proportional to the effective Hilbert-space dimension.

\section{NOON-State Synthesis}
A natural target state to characterize entangled-state synthesis algorithms is the so-called NOON state
\begin{equation}
|\psi_{\subs{target}}\rangle = |0\rangle \otimes \frac{1}{\sqrt{2}} \left( |N, 0\rangle + |0,N\rangle \right),
\end{equation}
an entangled superposition of resonator states in which $N$ photons are in mode $A$ or mode $B$.  This state can be considered a generalization of the Bell and GHZ states, and has potential applications in quantum metrology \cite{Dowling08}.  An initial experiment to generate a `high' NOON state (with $N=3$) was performed using a particular preparation method \cite{Wang11, Merkel10}.  That method uses a pair of three-level systems to couple the resonators, and a state-selective swap which limits the coupling rate due to the anharmonicity of these system \cite{Strauch11} (refinements of this method \cite{Su2014, Xiong2014} also use state-selective swaps).  Comparison of that approach with the original state-synthesis algorithm \cite{Strauch2010} (the photon subtraction method presented above) showed that both were experimentally comparable as far as decoherence is concerned \cite{Strauch2012b}.  Here we consider the new state-synthesis algorithm presented here and show that no state-selective interactions are required, allowing for faster operations and simplified experimental design.  Furthermore, using this photon swapping algorithm any state of the form
\begin{equation}
|\psi_{\subs{target}}\rangle = |0\rangle \otimes \sum_{n=0}^N c_n |N-n,n\rangle 
\label{diagonal_state}
\end{equation}
can be synthesized without state-selective interactions.

\begin{figure}
\begin{center}
\includegraphics[width=5.5in]{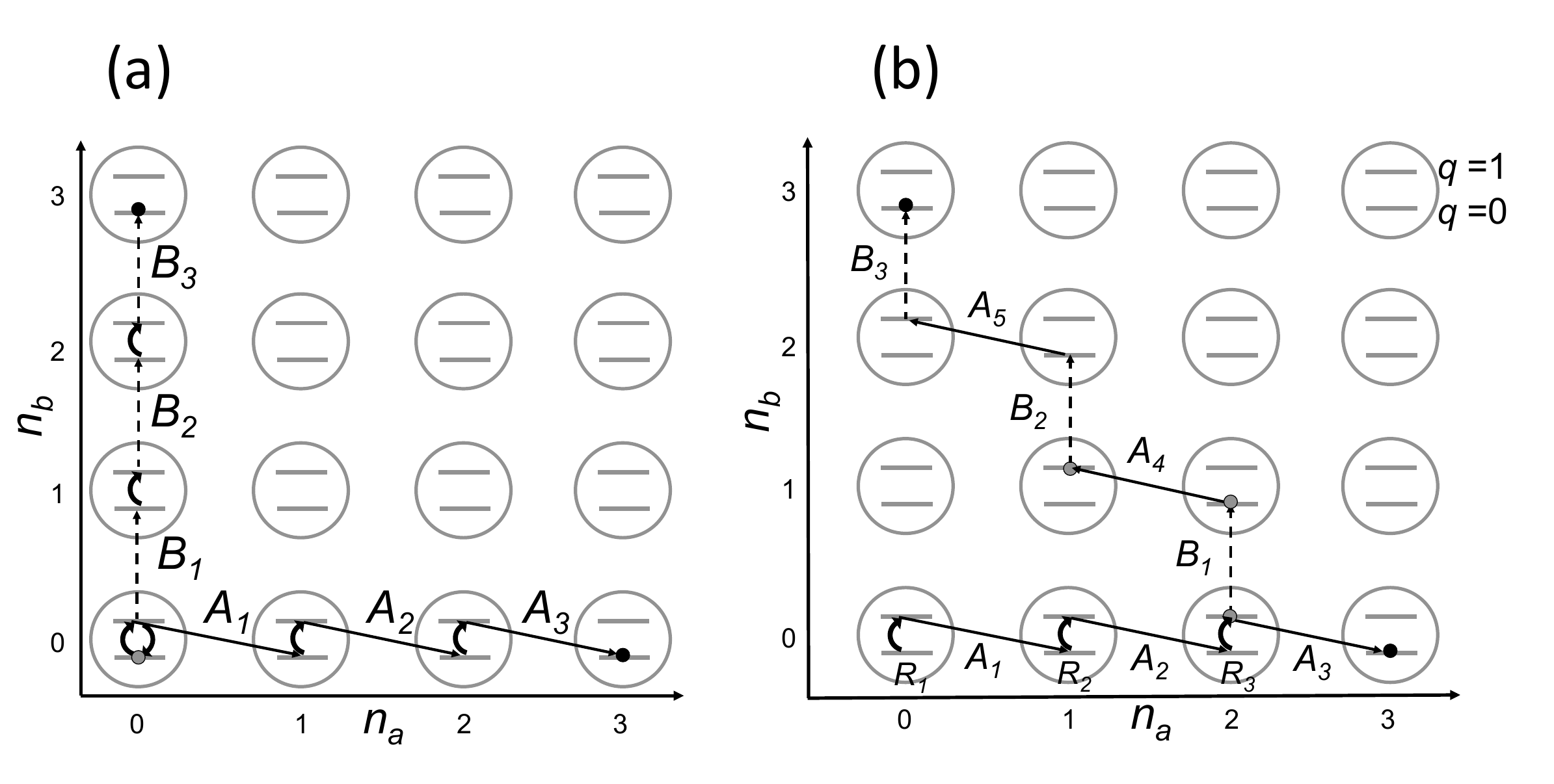}
\caption{NOON state synthesis (with $N=3$) using (a) the photon subtraction algorithm, and (b) the photon swapping algorithm.  Both prepare the NOON state with a linear number of steps, but the latter uses no state-selective interactions.}
\label{noon_fig}
\end{center}
\end{figure}

This improved performance is due to the paths through the Fock-space diagram taken by the new algorithm.   By following the photon swapping method, starting from a superposition of a given diagonal $n_a + n_b = N$, the first time through one can move all of the population down to $|0,N-1,0\rangle$.  Thus, one need only use a Law-Eberly sequence along the path $n_b = 0$ to remove the photons from the system.  The result is that any ``diagonal'' state of the form Eq. (\ref{diagonal_state}) can be synthesized by one sequence of photon swaps followed by a Law-Eberly sequence with no state-selective interactions.  The specific set of parameters for NOON state synthesis with $N=3$ are shown for the photon subtraction and swapping algorithms are shown in Tables \ref{fockprogram1} and \ref{fockprogram2}, respectively, and graphically represented in Fig. \ref{noon_fig}.  

\begin{table}[ht] 
\caption{NOON State Synthesis by the Photon Subtraction Algorithm} 
\centering      
\begin{tabular}{l l r}  
\hline\hline                        
Step & Parameters & Quantum State \\ [0.5ex] 
\hline                    
$R_{1} $ & $\gamma_1= \pi/2, n_a = 0, n_b = 0$ & $|0,0,0\rangle - i |1,0,0\rangle$  \\ 
$A_{1} $ & $\theta_1= \pi/2$ & $|0,0,0\rangle -  |0,1,0\rangle$  \\ 
$R_{2} $ & $\gamma_2, n_a = 1, n_b = 0 = \pi$ & $|0,0,0\rangle + i |1,1,0\rangle$  \\ 
$A_{2} $ & $\theta_2= \pi/2\sqrt{2}$ & $|0,0,0\rangle +  |0,2,0\rangle$  \\ 
$R_{3} $ & $\gamma_3= \pi, n_a = 2, n_b = 0$ & $|0,0,0\rangle - i |1,2,0\rangle$  \\ 
$A_{3} $ & $\theta_3= \pi/2\sqrt{3}$ & $|0,0,0\rangle -  |0,3,0\rangle$  \\ 
$R_{4} $ & $\gamma_4= \pi, n_a = 0, n_b = 0$ & $-i |1,0,0\rangle - |0,3,0\rangle$  \\ 
$B_{1} $ & $\theta_4= \pi/2$ &  $-|0,0,1\rangle -  |0,3,0\rangle$  \\ 
$R_{5} $ & $\gamma_5= \pi,  n_a = 0, n_b = 1$ & $i |1,0,1\rangle - |0,3,0\rangle$  \\ 
$B_{2} $ & $\theta_5= \pi/2\sqrt{2}$ & $|0,0,2\rangle -  |0,3,0\rangle$  \\ 
$R_{6} $ & $\gamma_6= \pi,  n_a = 0, n_b = 2$ & $-i |1,0,2\rangle - |0,3,0\rangle$  \\ 
$B_{3} $ & $\theta_3= \pi/2\sqrt{3}$ & $-|0,0,3\rangle -  |0,3,0\rangle$  \\ 
[1ex]       
\hline     
\end{tabular} 
\label{fockprogram1}  
\end{table} 

\begin{table}[ht] 
\caption{NOON State Synthesis by the Photon Swapping Algorithm} 
\centering      
\begin{tabular}{l l l}  
\hline\hline                        
Step & Parameters & Quantum State \\ [0.5ex] 
\hline                    
$R_{1} $ & $\gamma_1= \pi $ & $- i |1,0,0\rangle$  \\ 
$A_{1} $ & $\theta_1= \pi/2$ & $- |0,1,0\rangle$  \\ 
$R_{2} $ & $\gamma_2= \pi$ & $+ i |1,1,0\rangle$  \\ 
$A_{2} $ & $\theta_2= \pi/2\sqrt{2}$ & $ + |0,2,0\rangle$  \\ 
$R_{3} $ & $\gamma_3= \pi $ & $ - i |1,2,0\rangle$  \\ 
$A_{3} $ & $\theta_3= 0.2153$ & $-0.3643 |0,3,0\rangle - i 0.9313  |1,2,0\rangle$  \\ 
$B_{1} $ & $\eta_1= 2.1999$ &  $-0.3643 |0,3,0\rangle + i 0.548  |1,2,0\rangle - 0.753 |0,2,1\rangle $  \\ 
$A_{4} $ & $\theta_4= 1.3589$ & $+0.6454 |0,3,0\rangle -i 0.1283 |1,2,0\rangle + 0.2589 |0,2,1\rangle + i 0.7071 |1,1,1\rangle $  \\ 
$B_{2} $ & $\eta_2= \pi/2\sqrt{2}$ & $+0.6454 |0,3,0\rangle -i 0.2889 |1,2,0\rangle + 0.7071 |0,1,2\rangle  $  \\ 
$A_{5} $ & $\theta_5= \pi/2$ & $- 0.7071 |0,3,0\rangle - i 0.7071 |1,0,2\rangle$  \\ 
$B_{3} $ & $\eta_3= \pi/2\sqrt{3}$ & $- 0.7071 |0,3,0\rangle - 0.7071 |0,0,3\rangle$  \\ 
[1ex]       
\hline     
\end{tabular} 
\label{fockprogram2}  
\end{table} 

We now compare these two approaches for a general NOON state.  Based on previous analysis \cite{Strauch2012b}, we find that the photon subtraction method requires $N$ $A$-swaps, $N$ $B$-swaps, and $2N$ rotations.  No phase shifts are required, and the parameters scale as
\begin{eqnarray}
\left( \sum_n \langle \gamma_n \rangle \right)_{\subs{subtraction}} &\approx& \pi \left(2 N_{\subs{max}} - \frac{1}{2} \right), \nonumber \\
\left( \sum_n \langle \theta_n + \eta_n \rangle \right)_{\subs{subtraction}} &\approx& 6 \sqrt{N_{\subs{max}}} - 3.3.
\end{eqnarray}
As in the general photon subtraction algorithm, each of the rotations must be number-state-selective.  

The photon swapping algorithm requires $2N-1$ $A$-swaps, $N$ $B$-swaps, and $N$ rotations.  There are also a few phase shifts required, but they do no scale appreciably with $N$.  By looking at the numerical performance for the NOON state (not shown), we find
\begin{eqnarray}
\left( \sum_n \langle \gamma_n \rangle \right)_{\subs{swapping}} &\approx& \pi  N_{\subs{max}} \nonumber \\
\left( \sum_n \langle \theta_n + \eta_n \rangle \right)_{\subs{swapping}} &\approx& 9.9 \sqrt{N_{\subs{max}}} - 9.2
\end{eqnarray}
As described above, these rotations need not be number-state-selective.

Comparing these two algorithms for the NOON state, we see that we have traded rotations for swaps, with the photon swapping algorithm achieving the optimal number of rotations.  Even better, the photon swapping method does not require those rotations to be state-selective.  Thus, the method presented here has advantages both theoretically and experimentally, with the potential for fast performance and optimal scaling.

\section{Conclusion}
In this paper, we have studied state synthesis algorithms for superconducting resonators.  By reviewing the qudit and Law-Eberly schemes,  we have shown how solving for the inverse evolution allows one to determine the operations needed to synthesize an arbitrary state.  We have further shown how these step-by-step procedures have a complexity that typically grows linearly with the effective Hilbert-space dimension.  These schemes have been extended to two different state synthesis algorithms for a qubit coupled to two resonators.  The first type uses photon subtraction to ensure that the inverse evolution leads to the ground state, whereas the second uses photon swapping before any photons are removed from the system.  When taken in reverse, these algorithms allow one to synthesize an arbitrary entangled state of two resonators.   Finally, when applied to typical superconducting circuit experiments, we expect that the photon swapping method will have improved performance due to a reduced number of state-selective interactions.

While we have found an improved algorithm, we cannot claim to have found an optimal algorithm.  Indeed, we have reason to believe that numerical optimizations using the same basic Hamiltonians can lead to improved methods for state synthesis.  However, we also have reason to believe that the two algorithms compared here are the most natural analytical approaches to state synthesis.  At the same time, the differences between the two algorithms suggest that different types of optimizations may be possible.  The photon swapping algorithm minimizes the number of $A$-swaps performed on the system, but at the cost of a quadratic number of $B$-swaps and state-selective qubit rotations.  By constrast, the photon swapping algorithm minimizes the number of rotations, at the cost of an increased number of $A$-swaps and slightly increased overall complexity.  Nevertheless, for states such as the NOON state, the photon swapping method appears to have overall better performance, in that no state-selective rotations are needed at all.  

Finally, the {\em linear} scaling of the NOON state sequences are nearly ideal, in that the energy of the final state and the number of qubit rotations used (to put energy into the system) are both linear in the state number $N_{\subs{max}}$ \cite{Strauch11}.  We further observe that one can achieve a reduction in time complexity by a factor of two by driving multiple transitions simultaneously \cite{Strauch2012}, but the linear scaling remains.   However, recent work has found, using numerical optimization, that {\em sublinear} scaling is possible for Fock state preparation by starting from a large-amplitude coherent state and optimized displacements of the resonator \cite{Krastanov2015}; extending such a scheme to NOON state synthesis is an interesting question.  

In conclusion, we have improved the theoretical understanding and performance of entangled-state synthesis algorithms for superconducting resonators.  We hope that the results presented here, on a fundamental quantum control problem, may provide useful benchmarks for future explorations of control of superconducting or other resonator-based systems.   

\acknowledgements
This work was supported by the NSF under Project Nos. PHY-1005571 and PHY-1212413.

\bibliography{report}
\end{document}